\documentclass[10pt,a4paper]{article}

\usepackage[utf8]{inputenc}
\usepackage[T1]{fontenc}
\usepackage[margin = 2cm]{geometry}
\usepackage{setspace}
\usepackage{appendix}
\usepackage{amsmath}
\usepackage{amssymb}
\usepackage{mathtools}
\usepackage{graphicx}
\usepackage{amsmath}
\usepackage{listings}
\usepackage[export]{adjustbox}
\usepackage{subcaption}
\usepackage{enumitem}
\usepackage{tikz}
\usepackage[font={small,rm,it}]{caption}
\usepackage{float}
\usepackage{xcolor}
\usepackage{bm}
\usepackage{pdfpages}

\usepackage[style=numeric-comp,doi=false]{biblatex} 
\renewbibmacro{url+urldate}{}
\title{Supplementary Materials: Recoil-heating modulation in levitated optomechanics through optical wavefront shaping
}
\author{Hugo Marot$^1$, Joel Love Nanfack$^1$, Murad Abuzarli$^1$, Yann Louyer$^1$, Mathias Perrin$^1$ \\ and Nicolas Bachelard$^{1,\dagger}$
\\$^1$CNRS, Universit\'e de Bordeaux, LOMA, UMR5798, Talence, France. \\
$^\dagger$ nicolas.bachelard@u-bordeaux.fr
}

\begin{document}


# Recoil-heating modulation in levitated optomechanics through optical wavefront shaping

Hugo Marot[1], Joel Love Nanfack[1], Murad Abuzarli[1], Yann Louyer[1], Mathias Perrin[1] and Nicolas Bachelard[1*].

[1]CNRS, Université de Bordeaux, LOMA, UMR5798, Talence, France.

*email: nicolas.bachelard@u-bordeaux.fr



Abstract: Optical levitation relies on light forces to trap mesoscopic objects in a vacuum and assemble pristine oscillators, whose mechanical properties are commonly tamed through laser-intensity modulation—e.g., to provide leverage over the stiffness or engineer quantum states. When approaching the quantum regime at low pressure, the oscillator's dynamics is governed by recoil heating, which stems from the backaction of scattered photons and whose control is highly desirable. Here, we demonstrate that spatial degrees of freedom accessible within the trapping-light field can be harnessed to modulate recoil heating. Specifically, we introduce an optimization routine that spatially shapes the optical field's wavefront to either maximize or minimize the recoil-heating rate independently of other parameters like the mechanical stiffness. Importantly, this approach can serve to devise sequences of patterns providing smooth and deep modulations. Implementable into common experimental setups, this work leads the way to the dynamic modulation of optomechanical coupling in optical levitation.

Conceptualized by Bragynski and co-workers in the 70's [1], optomechanics investigates the interplay between light and mechanical modes at small scales. Over the years, levitated optomechanics has progressively emerged as major a contributor to the field thanks to a remarkable decoupling from surrounding environments [2]. As a specific example, rather than relying on manufactured devices [3], small-scale particles can be levitated in a vacuum using optical tweezers to assemble pristine resonators, whose mechanical properties are governed by trapping light and subsequent forces. As simple as they sound, and despite their moderate mechanical frequencies and small volumes, these macroscopic systems can reach extremely-high quality factors as they are free from any mechanical bound and only couple to their surrounding through electromagnetic field radiations [4]. Recently, their outstanding properties have been harnessed in various domains and notably in macroscopic quantum physics (e.g., mechanical ground-state cooling [5–8], quantum squeezing [9], single-atom-collision detection [10,11]). Nonetheless, much-anticipated prospects still lay ahead—such as the engineering of quantum sensors [12] and nonclassical superposition states [13] or the conception of dark-matter detectors [14]. One of the key advantages provided by optically-levitated particles with respect to manufactured resonators (e.g., suspended membrane [15]) stands in the dynamic controllability over their optomechanical properties. For instance, their mechanical frequency can be temporally modulated through trap-intensity modulation for non-classical states preparation [9] or expansion protocols [16]. Moreover, relying on wavefront-shaping techniques [17,18], we recently demonstrated that a similar control can be delivered through a spatial modulation of the light field enforcing a reorganization of optical forces [19].

In the quantum regime, optomechanical dynamics is governed by recoil heating, which stems from the backaction of scattered photons and materializes into stochastic transfers of optical momenta to the particle [20]. Specifically, at low pressure, thermal noise becomes negligible and recoil heating turns into the dominant driving mechanism, thus dictating quantum properties [21]. On the one hand, reducing recoil heating is often desirable as it generates decoherence channels and impinges the preparation of quantum states [22]. On the other hand, large recoil heating is intertwined with efficient detection schemes [23] as well as indicative of strong optomechanical coupling and quantum cooperativity [6]. Recently, different theoretical works have suggested various approaches to either reduce (e.g., reorganization of the optical-state density [24,25], injection of quantum-squeezed light [26]) or enhance recoil heating (e.g., levitation of highly-scattering particles [27]). More broadly in optomechanics, proposals have also emphasized that the ability to dynamically and independently modulate both mechanical frequency and optomechanical coupling (i.e., recoil heating), could offer mesmerizing opportunities for the engineering of nonclassical states [28,29]. Yet, to this day, no approach has been put forward to deliver a dynamic modulation of recoil heating. Following our recent experimental enhancements of mechanical resonances through wavefront shaping (see [19] and follow-up work [30]), the question can be asked whether this approach can be emulated to provide a dynamic control over recoil heating.

In the context of optically-levitated nanoparticles, here we theoretically demonstrate that the degrees of freedom accessible through the spatial shaping of the trapping field can be harnessed to modulate recoil heating. We first derive the impact of wavefront shaping over the optical forces exerted onto a nanoparticle as well as the inelastically-scattered field governing recoil heating. In particular, we report that scattering forces, which are usually neglected in optical levitation, do play a significant role. Then, using an optimization strategy, we show that recoil-heating maximization/minimization can be achieved, while preserving the mechanical frequency of the optical trap. Optimization parameters can be tuned to devise a sequence of patterns performing a continuous and dynamic recoil-heating modulation. At last, we discuss

the influence of particle and field parameters (e.g., radius, numerical aperture) onto the optimization’s performances. This work emphasizes the role of wavefront shaping to tame the quantum properties of optomechanical systems, while providing a never-reported controllability over the optomechanical interactions displayed in optically-levitated objects.

At first, we express how a phase-modulated beam effectively reshapes the field that is inelastically scattered by the oscillating particle (i.e., momentum transfer), thus providing leverage over recoil heating. We consider a dipolar particle of radius $r$, mass $m$ and complex polarization $\alpha = \alpha' + i\alpha''$. The particle is levitated in a vacuum using a microscope objective of focal lens $f$ and numerical aperture $NA$ (Figure 1**a** and [31]). Near the origin $[x, y, z] = [0,0,0]$, this objective focuses an $x$-polarized incoming field characterized by a wavelength $\lambda$ and a uniform amplitude $E_0$—thus forming an optomechanical oscillator of longitudinal frequency $\Omega_z$. A modulating device (not shown), such as a spatial light modulator [17,32], is located ahead of the objective and imprints onto the light field a phase pattern or wavefront $g(\theta, \phi)$—$\theta$ and $\phi$ standing for angular coordinates (Figure 1**b**). Our study is restricted to radial patterns (i.e., $g(\theta, \phi) = g(\theta)$) such that particles are trapped on-axis at a location $z_0$ where the focused field reads

$$\vec{E}_f(z_0) = \frac{i\pi f}{\lambda} e^{-ikf} E_0 (C_0 + iS_0)\vec{x}, \tag{1}$$

for which we introduced the coefficients $C_n$ and $S_n$ defined as

$$\begin{cases} C_n = \int_0^{\theta_{max}} \sqrt{\cos\theta} \sin\theta \, (1 + \cos\theta) \cos^{\mathrm{n}}\theta \cos(g(\theta) + 2\pi z_0/\lambda \cos\theta) \, d\theta \\ S_n = \int_0^{\theta_{max}} \sqrt{\cos\theta} \sin\theta \, (1 + \cos\theta) \cos^{\mathrm{n}}\theta \sin(g(\theta) + 2\pi z_0/\lambda \cos\theta) \, d\theta \end{cases} \tag{2}$$

From eqs. (1) and (2), we obtain, $\vec{E}_s(\theta, \phi; z_0)$, the field scattered towards the far-field angular coordinates $(\theta, \phi)$ by a particle located at $z_0$. For small displacements $\Delta z$, this field expands into $\vec{E}_s(\theta, \phi; z_0 + \Delta z) \approx \vec{E}_s(\theta, \phi; z_0) + \Delta z \partial_z \vec{E}_s(\theta, \phi; z)_{z_0}$, namely the sum of an elastic and an inelastic scattering term [33], respectively, whose expressions are provided in [31]. The distribution of inelastic scattering portrays the transfer of light momentum to the nanoparticle. Specifically, integrating the inelastic-scattering intensity over the whole far field provides access to the recoil-heating rate that reads

$$\Gamma = \frac{\beta E_0^2}{\Omega_z} \left( 2(C_0^2 + S_0^2) + 5(C_1^2 + S_1^2) \right), \tag{3}$$

in which $\beta$ is a constant defined in [31]. While the phase pattern, $g(\theta)$, only tunes the phase of the elastic-scattering distribution, eq. (3) emphasizes that it represents an efficient apparatus to spatially reshape the inelastic-scattering distribution and subsequently recoil heating. Importantly, we stress that this expression is consistent with former works (e.g., [23,34]) in the limit $S_n \to 0$—considering a uniform wavefront (i.e., $g(\theta) = 0$) and neglecting the mean contribution of scattering forces (i.e., trapping on focus, $z_0 = 0$).

We now demonstrate that, in line with our former experimental observations [19], mean scattering forces play a significant role in the trapping of nanoparticles and, in turn, impact the recoil-heating rate. Combining eqs. (1) and (2), we derive the mean longitudinal force, $\langle F_z \rangle$, which decomposes into a gradient, $\langle F_g \rangle$, and a scattering contribution, $\langle F_s \rangle$, fulfilling $\langle F_z \rangle = \langle F_g \rangle + \langle F_s \rangle$ (see caption in Figure 1 and [31]). Stable trapping at $z_0$ is achieved when both contributions balance off (i.e., $\langle F_z \rangle = 0$), which reduces to

$$\alpha'(S_0 C_1 - C_0 S_1) + \alpha''(C_0 C_1 + S_0 S_1) = 0, \tag{4}$$

while simultaneously requiring a positive stiffness $m\Omega_z^2 = -\partial_z \langle F_z \rangle|_{z_0} > 0$. As the optical theorem imposes $\alpha'' \neq 0$ [35], the second term in eq. (4) emphasizes that, in contrast with usual assumptions [33,34,36], $\langle F_s \rangle$ has a non-zero contribution that impacts optomechanical properties—e.g., $z_0$ and $\Omega_z$. In particular, for a uniform pattern it prevents trapping at $z_0 = 0$, where $\Gamma$ and, therefore, the optomechanical coupling would be optimal [31]. On a physical stand point, this highlights that the scattering force's fluctuations (encoded into $\Gamma$) relate to its mean value (i.e., $\langle F_s \rangle$). On a theoretical stand point, eqs. (3) and (4) stress that the influence of non-uniform wavefronts can be solely described using the coefficients $C_n$ and $S_n$ introduced in eq. (2).

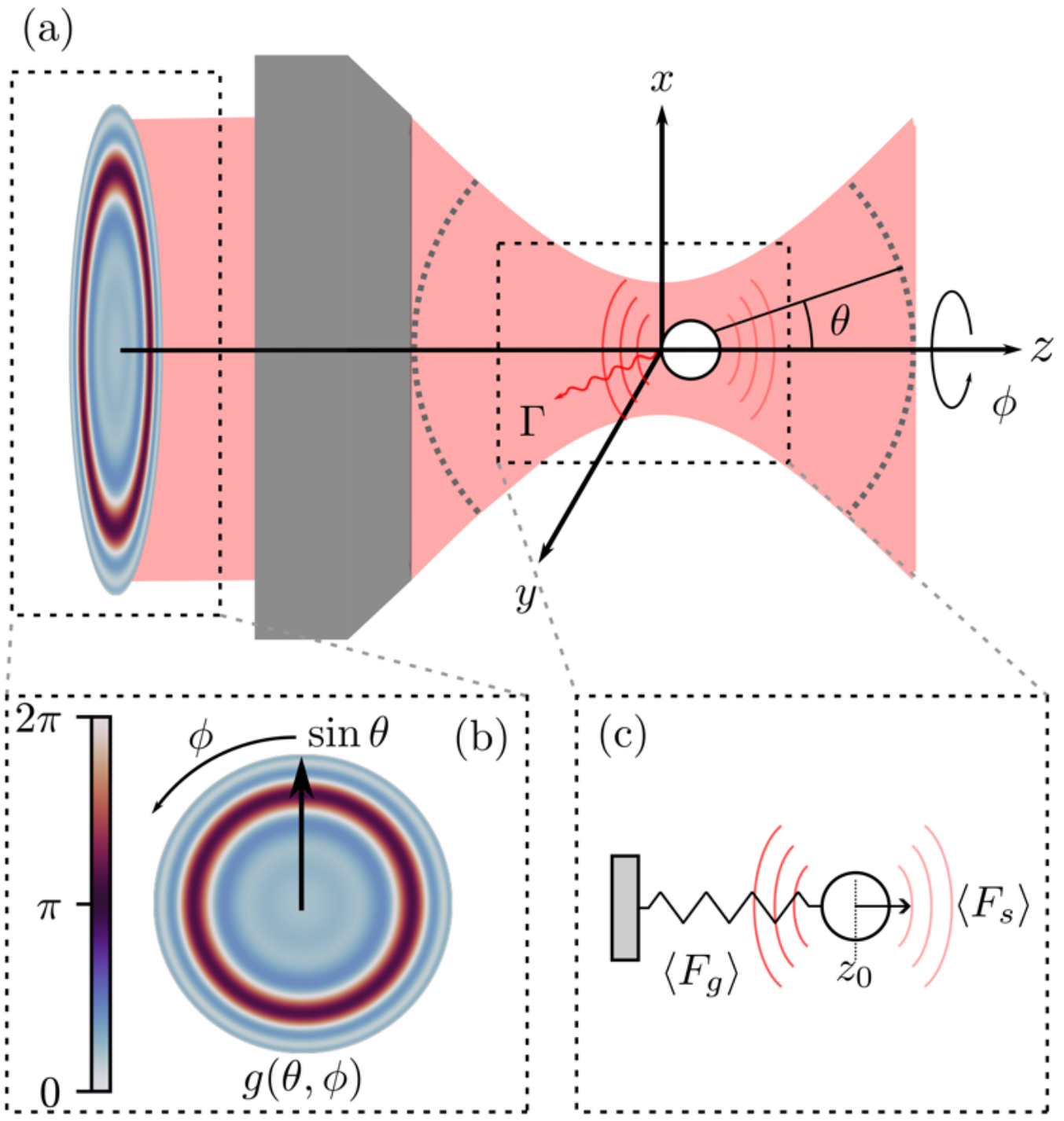


Figure 1 : **a**- In a vacuum, a microscope objective (grey) focuses near the origin $[x, y, z] = [0,0,0]$ a continuous light field (red beam, wavelength $\lambda$) to optically trap a dipolar particle (white sphere, radius $r$, mass $m$), which behaves as a mechanical oscillator (frequency $\Omega_z$). The angles $\theta$ and $\phi$ stand for far-field coordinates. The stochastic transfer of optical momentum induced by individual photons onto the particle results in an inelastically-scattered field (concentric semi-circles), whose overall contribution translate into recoil heating, $\Gamma$. Upstream from the objective, a modulating device (not shown) performs a spatial modulation of the light beam before focusing (blue-to-red concentric circles), which tunes the optical forces exerted onto the particle and ultimately $\Gamma$. **b**- The phase-modulation pattern (or wavefront) in **a** is described as a function $g$, whose radial and azimuthal coordinates read respectively $sin(\theta)$ and $\phi$. **c**- Mechanically, the mean optical force is decomposed into a conservative contribution known as the gradient force (akin to a spring), $\langle F_g \rangle$, and a non-conservative force known as the scattering force, $\langle F_s \rangle$. Both contributions balance off at the trapping location $z_0$ around which the particle oscillates.

We implement a routine that iteratively reshapes the trapping field in order to optimize the recoil-heating rate, while preserving the oscillator stiffness (i.e., constant $\Omega_z$). At each iteration, the wavefront $g(\theta)$ is expanded as a linear combination of Zernike polynomials, which is adjusted to perform either a minimization or a maximization of $\Gamma/\Omega_z$ [31]. Independent of optical power, this cost function is optimized under two constraints, imposing respectively stable trapping at $z_0$ (i.e., to fulfill eq. (4)) and a range over which $\Omega_z$ can be tuned. Finally, when the optimization has converged, $\Omega_z$ is brough back to its initial value through an adjustment of field intensity $E_0^2$ that also tunes the value of $\Gamma$. In Figure 2**a**, we provide an example of recoil-heating minimization, which converges towards the optimized pattern displayed in inset and characterized by distinctives phase rings (blue-to-red circles). Specifically, are displayed in thick black (respectively red) $\langle F_g\rangle(z)$ and in dashed black (respectively red) $-\langle F_s\rangle(z)$ that correspond to the landscapes obtained before (respectively after) the procedure, while their intersection is marked by a black tick (respectively red) corresponding to the equilibrium position $z_0$. If the slope of the gradient force close to $z_0$ (mainly responsible for the optical confinement) is barely modified, the scattering force has been largely reduced. When adjusting the optical power after the optimization to keep $\Omega_z$ to its initial value (and provide a fair comparison), the final rate, $\Gamma_{opt}$, is reduced by almost a factor 2 with respect to the initial one, $\Gamma_0$ ($\Gamma_{opt} = .56\,\Gamma_0$). Inside the black and red-dashed box of Figure 2**b**, are displayed the far-field inelastic-scattering intensity and phase distributions associated with the uniform and optimized case of Figure 2**a,** respectively. These plots emphasize that such a recoil-heating reduction originates from a deep reshaping of inelastic scattering. As shown in Figure 2**c**, the procedure can then be adjusted to provide a recoil-heating magnification ($\Gamma_{opt} = 1.60\,\Gamma_0$), which is characterized by an enhancement of the mean scattering force and the inelastic scattering distribution [31]. In sharp contrast with the minimization case of Figure 2**a**, the optimization pattern (inset) is now mainly modulated on its edge (i.e., large $\theta$) while remaining mostly unform on its center (i.e., low $\theta$). Such a discrepancy seems to be systematic and indicative that the algorithm tends to deal with higher (respectively lower) spatial frequencies associated with large $\theta$ (respectively lower $\theta$) to maximize (respectively minimize) the transfer of momentum through inelastic scattering and, thus, recoil-heating [31]. We stress that larger recoil-heating enhancements/reductions can be reached when allowing for larger variations of $\Omega_z$ [31]. Moreover, the use of continuous Zernike polynomials should ease the experimental implementation of the procedure. We also confirmed that our simulations are consistent with a numerical model relying on a convergent expansion onto electric and magnetic multipoles [37] and we checked the stability of the optimized trapping fields along the transverse directions $x$ and $y$ [31].

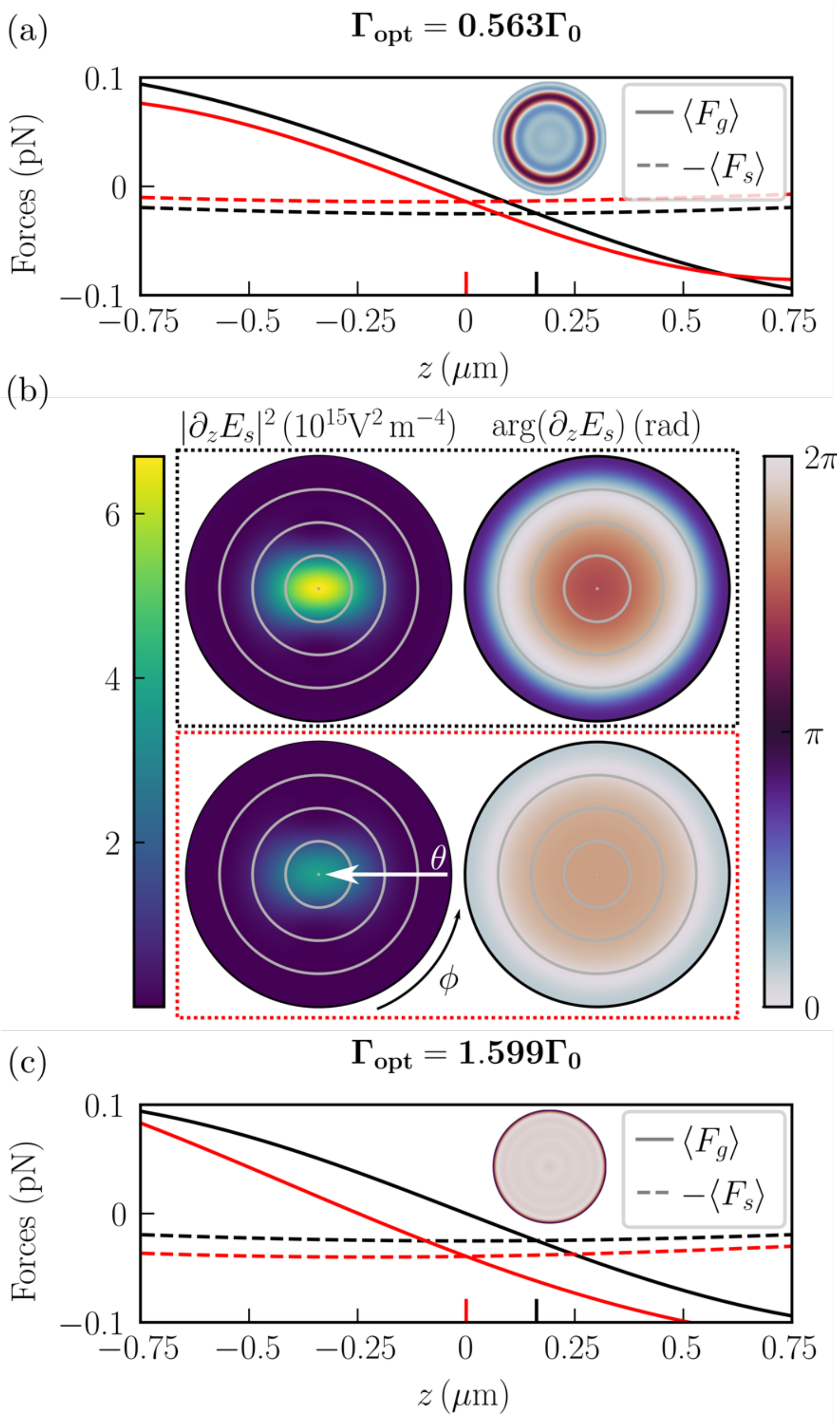


Figure 2 : **a**- Recoil-heating minimization for a 75 $nm$-radius nanoparticle illuminated by a 1064 $nm$ and 40 $mW$ light field, while using a $NA = 0.8$ objective. The landscapes of $\langle F_g \rangle$ (thick black curve) and $-\langle F_s \rangle$ (dashed black curve) are obtained under a uniform phase pattern and cross at the trapping location, $z_0$ (black tick). The landscapes of $\langle F_g \rangle$ (thick red curve) and $-\langle F_s \rangle$ (dashed red curve) are obtained under the wavefront displayed in inset (blue-to-red concentic circles) and cross at the trapping location, $z_0$ (red tick). **b**- Top left (respectively right) diagramm inside the dashed black frame. Inelastically-scattered intensity (respectively phase) radiated over the whole far field (i.e., $\theta \in [0, \pi]$ and $\phi \in [0,2\pi]$), which is obtained under a uniform phase pattern. Bottom left (respectively right) diagramm inside the dashed red frame. Inelastically-scattered intensity (respectively phase) radiated over the whole far field (i.e., $\theta \in [0, \pi]$ and $\phi \in [0,2\pi]$), which is obtained using the pattern displayed in **a**. **c**- Recoil-heating maximization using the parameters considered in **a** (same color code). The optimized wavefront is provided in inset. In both panels **a** and **c**, the trapping location after optimization is set to $z_0 = 0$ as it can be translated at will by adding a defocus to $g(\theta)$ [31].

Now, we harness the optimization scheme to assemble a series of phase patterns achieving a modulation of $\Gamma$ independently of the mechanical frequency $\Omega_z$. Switching amongst these patterns over time would effectively provide a temporal modulation of the optomechanical coupling. As explained in [31], adjusting the range over which $\Omega_z$ can be tuned enables the algorithm to converge towards different wavefronts providing specific maximized/minimized recoil-heating rates. In Figure 3**a**, we plot in red the evolution of $\Gamma$ when switching across a series of 12 patterns optimized to reach distinct rates—the curve is fitted by a sinusoidal-like curve in dashed grey. The evolution across the series is labelled using a time-like variable $\tilde{t}$ with arbitrary units. Five of these patterns are provided in inset. They confirm that strong enhancements rely on high-frequency components (e.g., $\tilde{t} = 0.14$ and $0.2025$), while weak modulations are provided by almost uniform patterns (e.g., $\tilde{t} = 0.265$) and significant reductions by low-frequency ones (e.g., $\tilde{t} = 0.3275$ and 0.39). Remarkably, using here commonly used experimental parameters, we report a peat-to-peak modulation depth of $\approx 65\%$. In Figure 3**b**, we plot the intensity adjustment over time ($E_0^2$, blue curve) that is required to enforce a steady mechanical frequency $\Omega_z$ (black curve). It oscillates twice faster than the curve in Figure 3**a** as both minimizing and maximizing may require the same intensity adjustment. If each pattern is effectively associated with a different $z_0$, we explain in [31] that the patterns can be adjusted to enforce a steady trapping location all throughout the modulation.

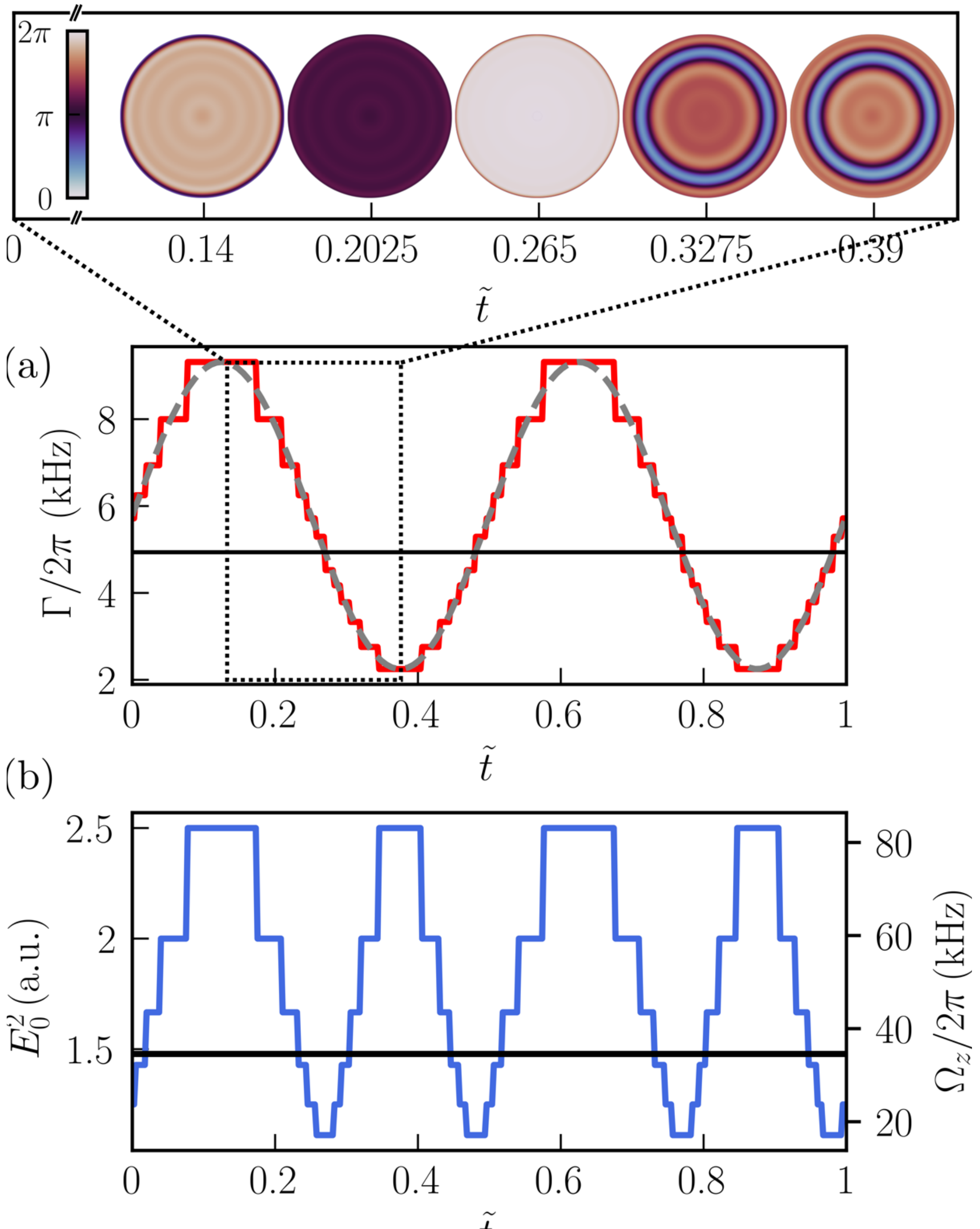


Figure 3 : **a**- Periodic recoil-heating evolution achieved by successively alterning amongst a series of 12 optimization patterns (red curve). The black line marks the rate obtained for a uniform wavefront, $\Gamma_0$. A sinusoidal-like fit is provided in dashed grey, while the time scale $\tilde{t}$ represents arbitrary values. The inset above displays five patterns used throughout the modulation (dashed frame). The two left-most patterns ($t = 0.14$ and $t = 0.2025$) produce recoil-heating enhancements, while the one in the middle ($t = .265$) achieves barely no modulation and the two right-most ($t = 0.3275$ and $t = 0.39$) provide reductions. **b**- Intensity modulation $E_0^2$ achieved over time (blue curve), which holds the resonance frequency $\Omega_z$ constant (black line).

At last, we show that recoil-heating modulation can be achieved over a large range of particles' sizes, while being unaffected by the objective's numerical aperture. In Figure 4**a**, we vary the nanoparticle's radius within the range of dipolar scattering (i.e., $kr < 1$). For each radius, $r$, we mark in black the initial rate, $\Gamma_0$, while the two red crosses mark the maximized and minimized rates, $\Gamma_{opt}$, that are obtained following the approach described in Figure 2. If as expected the rate increases with radius (as particle's scattering strengthen), we observe that the modulation depth remains almost identical irrespectively of the size. In Figure 4**b**, we then vary the numerical aperture of the trapping objective, which manifests in eq. (2) and therefore deeply influences all optomechanical properties. Using a similar color code as in Figure 4**a**, we report that optimization performances should remain almost identical over the range of numerical apertures used in common optical-levitation experiments [2].

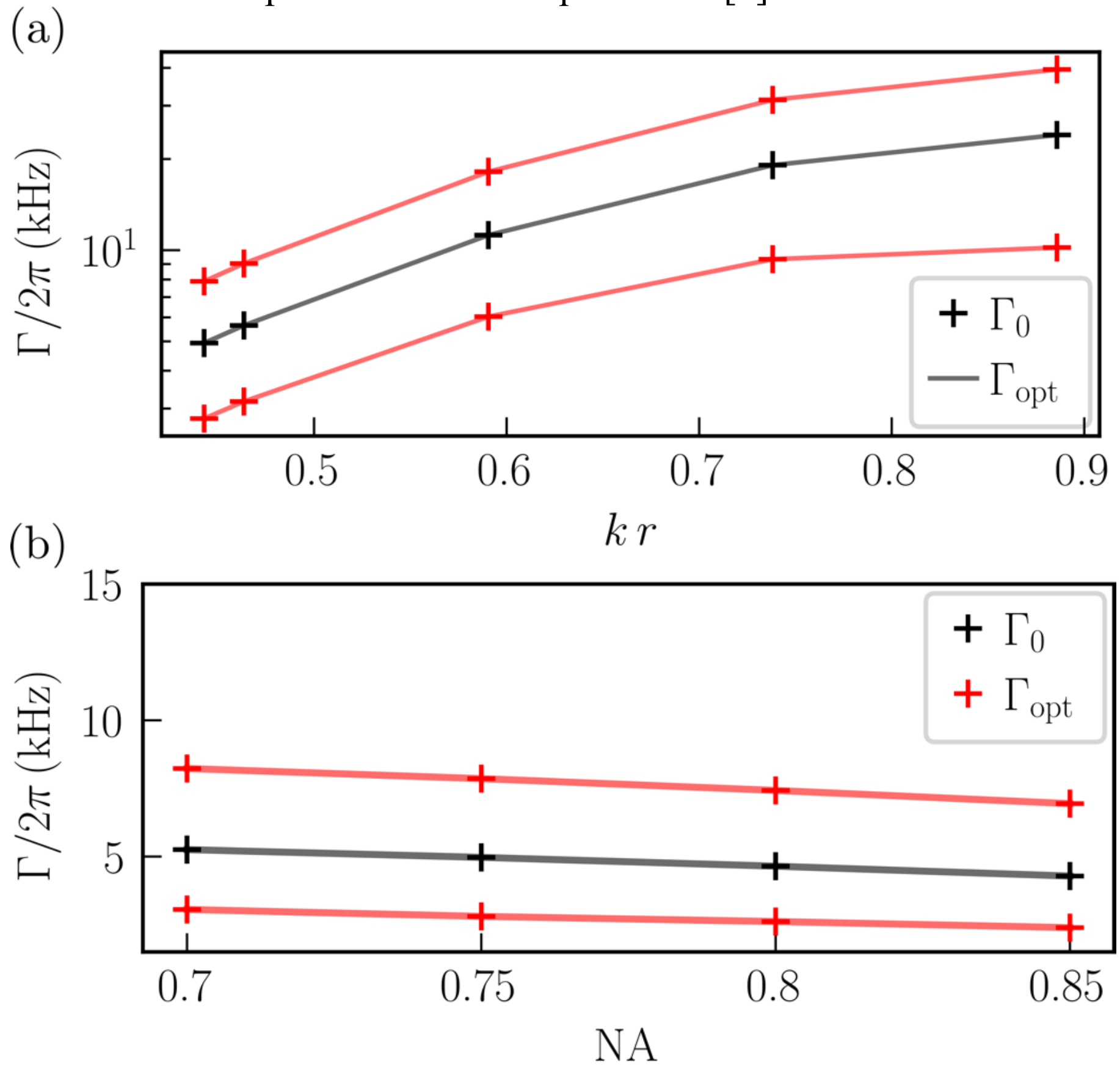


Figure 4 : **a**- The maximization and minimization provided in Figure 2 are reproduced for different particles' radii. The black crosses represent the rate $\Gamma_0$ obtained under a uniform pattern, while the red crosses represent the maximal and minimal values $\Gamma_{opt}$ reachable under the constraints used in Figure 2. **b**- Recoil-heating maximization and minimization for a 75 $nm$-radius particle, while varying the objective's numerical aperture. Same color code as in **a**.

In conclusion, here we demonstrated that the recoil heating exerted onto a levitated particle can be either reduced or enhanced through a spatial shaping of the trapping beam. We also showed that this wavefront-shaping approach can be adapted to provide a strong and almost-continuous recoil-heating modulation. Moreover, we highlighted that such a modulation is fully compatible with the experimental requirements of current levitation setups, while benefiting from a large operating range. Overall, this work emphasizes that spatial control over the trapping beam is provided with multiple degrees of freedom [32,38], which are still vastly unexplored and could serve to tame others optomechanical features (e.g., detection efficiency [23]). Importantly, this approach could be readily implemented into current setups and deliver straightforward improvements (e.g., more efficient ground state cooling [6], enhanced ponderomotive squeezing [39,40]). With the current development of ultra-high-speed spatial light modulators [41], this work opens the door to the implementation of control schemes relying on the resonant modulations of optomechanical interactions and offering to deliver exotic quantum mechanical states [28,29]. Aside from optomechanics, these results could also reveal impactful to engineer non-equilibrium steady states through time protocols—such as shortcuts to adiabaticity [42] or Floquet modulations [43].



\maketitle
\tableofcontents

\appendix
\newpage
\section{Derivation of the recoil-heating rate}

\begin{figure}[h]
    \centering
    \includegraphics[width=0.75\linewidth]{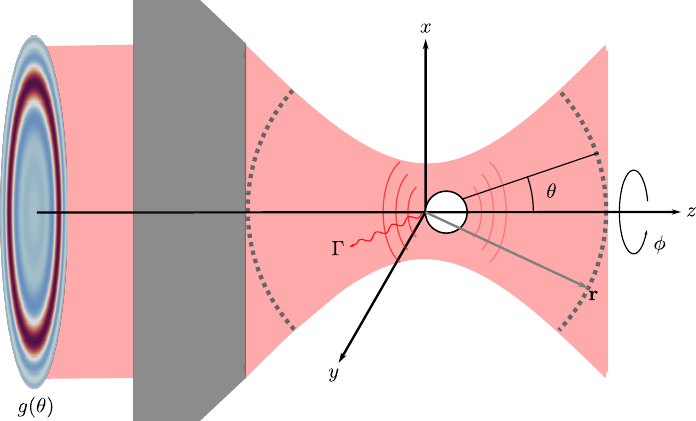}
    \caption{A nanoparticle akin to an electric dipole (white sphere) is trapped in a vacuum near the focal spot of a high-NA objective of focal length $f$.  
    We note $[x,y,z]$ the cartesian coordinates, whose origin coincides with the objective's focus while the particle is trapped along the optical axis at a location $[0,0,z_0]$.
    We note $(\theta,\phi)$ the angular coordinates pointing towards the far field (gray dots). 
    Specifically, a far-field location $\textbf{r}$ reads in spherical coordinates $\textbf{r}=[f,\theta,\phi]$.
    The incoming CW light field (red beam) is characterized by a wavelength $\lambda$, while its phase profile is spatially modulated by a dedicated device (e.g., SLM) that imprints a non-uniform wavefront $g(\theta)$ (blue-to-red concentric circles).  
    The light scattered by the particle (concentric red arc-circles) is responsible for recoil heating that materializes through the rate $\Gamma$.}
    \label{fig:Scheme}
\end{figure}

First, we compute the electric field scattered by a dipolar particle of polarizability $\alpha = \alpha'+i\alpha''$, which is trapped near the focus of an objective of numerical aperture $\text{NA}$. 
When considering an axial symmetric pattern $g(\theta)$, the field created by an infinitely overfilled microscope objective near its focus reads \cite{Novotny_Hecht_2012_4}:
\begin{equation}
\label{Full_field}
\begin{split}
        \mathbf{E}_f(\rho, \phi, z) = \frac{i\pi f}{\lambda}e^{-ikf}E_0 \begin{bmatrix}
            I_{00} + I_{02}\cos(2\phi) \\ I_{02}\sin(2\phi) \\ -2iI_{01}\cos(\phi)
        \end{bmatrix}_{[x, y, z]},
    \end{split}
\end{equation}
where the incoming light is $\textbf{x}$-polarized, $\rho = \sqrt{x^2+y^2}$ and $k=2\pi / \lambda$. 
The integrals \(I_{00}\), \(I_{01}\) and \(I_{02}\) are given in spherical coordinates by:
\begin{equation}
\label{I_Integrals}
    \begin{split}
        &I_{00} = \int_0^{\theta_\mathrm{max}}\sqrt{\cos\theta}(1+\cos\theta)\sin\theta J_0\left(\frac{2\pi\rho}{\lambda}\sin\theta\right)e^{ig(\theta) + i\frac{2\pi z}{\lambda}\cos\theta} d\theta, \\
        &I_{01} = \int_0^{\theta_\mathrm{max}}\sqrt{\cos\theta}\sin^2\theta J_1\left(\frac{2\pi\rho}{\lambda}\sin\theta\right)e^{ig(\theta) + i\frac{2\pi z}{\lambda}\cos\theta} d\theta, \\
        &I_{02} = \int_0^{\theta_\mathrm{max}}\sqrt{\cos\theta}(1-\cos\theta)\sin\theta J_2\left(\frac{2\pi\rho}{\lambda}\sin\theta\right)e^{ig(\theta) + i\frac{2\pi z}{\lambda}\cos\theta} d\theta,
    \end{split}
\end{equation}
where \(J_n\) stands for the Bessel function of the first kind for \(n \in \mathbb{N}\). 
As the wavefront $g(\theta)$ is axis-symmetric, mean scattering forces are null in the transverse directions and stable trapping will occur along the optical axis (i.e., \(\rho = 0\)). 
Thus, trapping at the location $z=z_0$  is only governed by $I_{00}$, which simplifies as: 
\begin{equation}
\label{I_00}
    I_{00} = \int_0^{\theta_\mathrm{max}}\sqrt{\cos\theta}(1+\cos\theta)\sin\theta e^{ig(\theta) + i\frac{2\pi z_0}{\lambda}\cos\theta} d\theta.
\end{equation}
Thus, the trapping field near the trapping position \([0, 0, z_0]\) is akin to an \(\textbf{x}\)-polarized plane wave: 
\begin{equation}
\label{Simplified_field}
    \mathbf{E}_f(z_0)  = \frac{i\pi f}{\lambda}e^{-ikf}E_0(C_0+iS_0)\textbf{x}, 
\end{equation}
where \(C_0\) and \(S_0\) are the real and imaginary parts of \(I_{00}\), respectively. 

More generally, we introduce for all \(n \in \mathbb{N}\) the coefficients:
\begin{equation}
\label{CS}
    \begin{split}
        &C_n = \int_0^{\theta_\mathrm{max}}\sqrt{\cos\theta}(1+\cos\theta)\cos^n\theta\cos\left(g(\theta)+\frac{2\pi z}{\lambda}\cos\theta\right)d\theta, \\
        &S_n = \int_0^{\theta_\mathrm{max}}\sqrt{\cos\theta}(1+\cos\theta)\cos^n\theta\sin\left(g(\theta)+\frac{2\pi z}{\lambda}\cos\theta\right)d\theta,
    \end{split}
\end{equation}
fulfilling: \(\partial_zC_n = -(2\pi/\lambda) S_{n+1}\) and \(\partial_zS_n = (2\pi/\lambda) C_{n+1}\).\\

Next, we derive the field scattered by the dipolar particle. 
The field scattered towards a far-field position \(\mathbf{r}\) by a dipole radiating at position \(\mathbf{r}_0\) is given by:
\begin{equation}
\label{Grenn_tensor}
    \mathbf{E}_s(\mathbf{r}) = \frac{k^2}{\epsilon_0\epsilon}G(\mathbf{r}, \mathbf{r}_0)\alpha\mathbf{E}(\mathbf{r}_0), 
\end{equation}
where $\epsilon$ and $\epsilon_0$ define respectively the material and vacuum permittivity, while \(G(\mathbf{r}, \mathbf{r}_0)\) stands for the far-field Green tensor \cite{Novotny_Hecht_2012_2}. 
In the case of spherical coordinates (as considered here) and for a particle located at $\textbf{r}_0=[0,0,z_0]$, the scattering field reads as follows: 
\begin{equation}
\label{Full_scattered_field}
    \begin{split}
        \mathbf{E}_s(\mathbf{r}) = \frac{ikf}{2}E_0e^{-ikf}(C_0+iS_0)\frac{\alpha'k^2}{4\pi\epsilon_0\epsilon}\frac{e^{i\mathbf{k}(\mathbf{r} - \mathbf{r}_0)}}{|\mathbf{r}-\mathbf{r}_0|}\begin{bmatrix}
            0 \\ \cos\theta\cos\phi \\ -\sin\phi
        \end{bmatrix}_{[r,\theta,\phi]}.
    \end{split}
\end{equation}
In the far field (gray-dotted curve, Figure \ref{fig:Scheme}), \(|\mathbf{r}| = f\), \(kr_0 = 2\pi z_0/\lambda \gg r_0/f = z_0/f\) and thus we find that:
\begin{equation}
\label{Farfield_scattered_field}
    \begin{split}
        \mathbf{E}_s(\mathbf{r}) = i \frac{E_0k^3\alpha'}{8\pi\epsilon_0\epsilon}(C_0+iS_0)e^{-ikz\cos\theta}\begin{bmatrix}
            0 \\ \cos\theta\cos\phi \\ -\sin\phi
            \end{bmatrix}_{[r, \theta, \phi]}.
    \end{split}
\end{equation}
\color{black}

For small displacements \(\Delta z\) around the trapping position $z_0$, the scattered field can be expanded at first order as \(\mathbf{E}_s(\theta, \phi, z_0 + \Delta z) \approx \mathbf{E}_s(\theta, \phi, z_0) + \left.\Delta z \partial_z\mathbf{E}_s(\theta, \phi, z)\right|_{z_0} \) \cite{Wang:25}. 
When the dipolar particle experiences position fluctuations \(\Delta z\), it re-emits photons through Stokes scattering and exchanges momentum with the light field. 
Thus, in the relation above, we say that the second term represents the field that is inelastically scattered by the particle at \(z_0\). 
Conversely, if the scatterer has no position fluctuations, it cannot exchange momentum with the field, so we say that the first term represents the field that is elastically scattered by the particle at \(z_0\). 
For a quantum harmonic oscillator in its ground state, displacement fluctuations relate to the zero point fluctuation: \(z_\mathrm{zpf} = \sqrt{\frac{\hbar}{2m\Omega_z}}\). 
The number of inelastically-scattered photons is given by the integral over the surface of the time-averaged Poynting vector, divided by the energy of a photon. 
In the far field, this reduces to: 
\begin{equation}
\label{Backaction}
    \Gamma = \frac{c\epsilon_0}{2\hbar\omega}\int_\Omega |z_\mathrm{zpf}\partial_zE_s(z)_{z_0}|^2 r^2d\Omega = \frac{8}{15}\frac{f^2\alpha'^2k^7}{256\pi\epsilon_0\epsilon^2}\frac{E_0^2}{m\Omega_z}(2(C_0^2+S_0^2)+5(C_1^2+S_1^2)).
\end{equation}
\begin{figure}[h]
    \centering
    \includegraphics[width=0.6\linewidth]{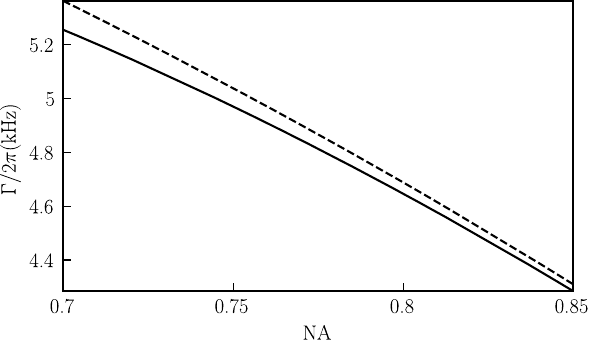}
    \caption{Recoil-heating rate, $\Gamma$, as a function of the numerical aperture, $\text{NA}$, in the case of a trap at focus (dashed line, eq. \eqref{NovotnyCoherence}) and in the case of a particle at $z_0$ (thick line, eq. \eqref{Backaction}). 
    Numerically, we considered a $\text{NA}$=0.8, a particle of radius \(r = 75\) nm and a light beam of wavelength \(\lambda = 1064\) nm and power $40$ mW. 
    As the recoil-heating rate is inversely proportional to the pulsation \(\Omega_z\), it is computed explicitly using trapping parameters and relevant integral coefficients. }
    \label{fig:NovotnyComparisson}
\end{figure}
\newpage
In the limit where \(z_{0} \to 0\), we recover the results from \cite{Tebbenjohanns2019}, stating that :
\begin{equation}
\label{Novotny_backaction}
    \Gamma = \frac{1}{2m\hbar\Omega_z}\left(\frac{2}{5} + A^2\right)\frac{\hbar k}{c}P_\mathrm{dip},
\end{equation}
in which \(P_\mathrm{dip}\) is the total power scattered by the dipole and \(A = \frac{C_1(z=0)}{C_0(z=0)}\). 
Computing the total scattered power, we find that:
\begin{equation}
\label{NovotnyCoherence}
    \Gamma = \frac{8}{15}\frac{f^2\alpha'^2k^7}{256\pi\epsilon_0\epsilon^2}\frac{E_0^2}{m\Omega_z}(2C_0^2+5C_1^2),
\end{equation}
which matches eq. \eqref{Backaction} when \(S_n = 0\) and \(z_0 = 0\).
In Figure \ref{fig:NovotnyComparisson}, the dashed curve plots the recoil-heating rate $\Gamma$, as the function of the objective's $\text{NA}$, for a particle at focus (i.e., given by eq. \eqref{NovotnyCoherence}), while the thick curve  consider a trap at \(z_0\) (i.e., given by eq. \eqref{Backaction}). 
We see that neglecting the particle equilibrium position leads to a small overestimation of the recoil-heating rate. 
As the numerical aperture increases, the equilibrium position gets closer to the focus (see Figure \ref{fig:eqpos}), thus explaining why the discrepancy between the two methods is reduced. 
Finally, we stress that the frequencies $\Omega_z$ used in eqs. \eqref{Backaction} and \eqref{Novotny_backaction} differ as the particle is not considered to be at the same location and thus not submitted to the same forces. 
Their respective expressions are provided in eqs. \eqref{Stiffness} and \eqref{StiffnessNovotny} derived in section \ref{section:Forces}.

\newpage
\section{Forces and equilibrium condition}
\label{section:Forces}
Now, we derive the gradient and the scattering force when applying a non-uniform wavefront $g(\theta)$. 
We then show that, although the scattering force is much smaller than the gradient force, it has appreciable effects on the trapping position, $z_0$. 
This, in turn, affects quantities such as the trap frequency, $\Omega_z$, or the recoil-heating rate, $\Gamma$.
For a dipole, the gradient force is derived through the gradient of the field intensity. 
Recalling the value of the field in eq. \eqref{Simplified_field}, we show that the force along the \(z\) direction is given by:
\begin{equation}
\label{Gradient_force}
    F_g(z_0) = \frac{\pi^2f^2}{\lambda^2}E_0^2\frac{\alpha'}{4}\partial_{z_0}|C_0+iS_0|^2 = \frac{\pi^3f^2E_0^2\alpha'}{\lambda^3}(S_0C_1-C_0S_1).
\end{equation}
The scattering force evolves with the gradient of the field phase and reads:
\begin{equation}
\label{Scattering_force}
    F_s(z_0) = \frac{\pi^2f^2}{\lambda^2}E_0^2(C_0^2+S_0^2)\frac{\alpha''}{2}\partial_{z_0}\arg(ie^{-ikf}(C_0+iS_0)) = \frac{\pi^3f^2E_0^2\alpha''}{\lambda^3}(C_0C_1+S_0S_1),
\end{equation}
while the trapping stiffness along \(z\) relates to both forces as follows:
\begin{equation}
\label{Stiffness}
    \kappa_z = m \Omega_z^2 = -\partial_{z_0}(F_g+F_s) = \frac{2\pi^4f^2E_0^2}{\lambda^4}[\alpha'(C_0C_2+S_0S_2-C_1^2-S_1^2) + \alpha''(C_0S_2-S_0C_2)].
\end{equation}
In the limit $z_0\rightarrow 0$, this stiffness reduces to:
\begin{equation}
\label{StiffnessNovotny}
    \kappa_z(z_0\rightarrow 0) = \frac{2\pi^4f^2E_0^2}{\lambda^4}\alpha'[C_0C_2-C_1^2].
\end{equation}

\begin{figure}[h]
    \centering
    \includegraphics[width=0.75\linewidth]{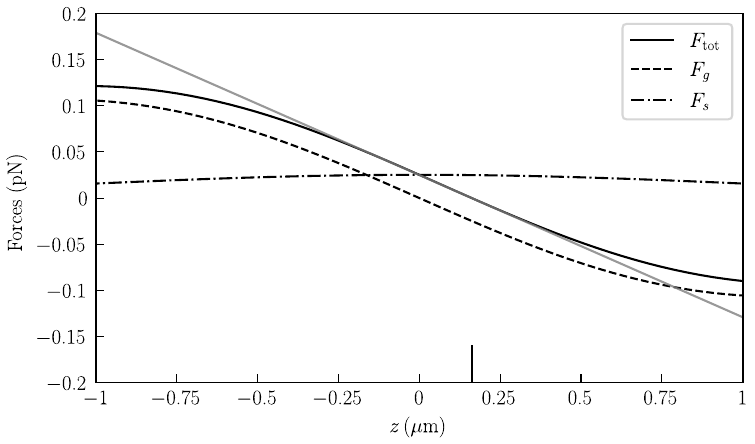}
    \caption{Longitudinal optical forces exerted by a uniform wavefront $g(\theta)=0$ onto a \( = 75\) nm-radius particle at position \(z\) (black), using an objective of $\text{NA}$=0.8 and a laser of wavelength $\lambda=1064$ nm and power of 40 mW.
    The dashed and point-dashed curves display respectively the gradient, $F_g$ and the scattering force, $F_s$. 
    The thick curve corresponds to the total force $F_{tot}=F_g+F_s$, which becomes zero at the trapping location $z_0$ (black tick).
    The slope of $F_{tot}$ at $z_0$ (corresponding to $-\kappa_z$) is provided by the light-gray thick line.}
    \label{fig:Stiffplot}
\end{figure}

In Figure \ref{fig:Stiffplot}, we plot the different longitudinal forces (i.e., along $z$), which are applied to the particle when its location $z$ is varied.
Stable trapping is enforced at the location $z_0$ (black tick) where the total force fulfills $F_{tot}=F_g+F_s=0$ (black curve), while the stiffness, $\kappa_z$, remains positive (gray line). 
As such, the condition for a stable equilibrium position can be expressed as a function of the coefficients $C_n$ and $S_n$ and boils down to:
\begin{equation}
\label{eq_position}
    S_0C_1-C_0S_1 + \frac{\alpha''}{\alpha'}(C_0C_1+S_0S_1) = 0 \quad \& \quad \alpha'(C_0C_2+S_0S_2-C_1^2-S_1^2) + \alpha''(C_0S_2-S_0C_2) >0
\end{equation}

\begin{figure}
    \centering
    \includegraphics[width=\linewidth]{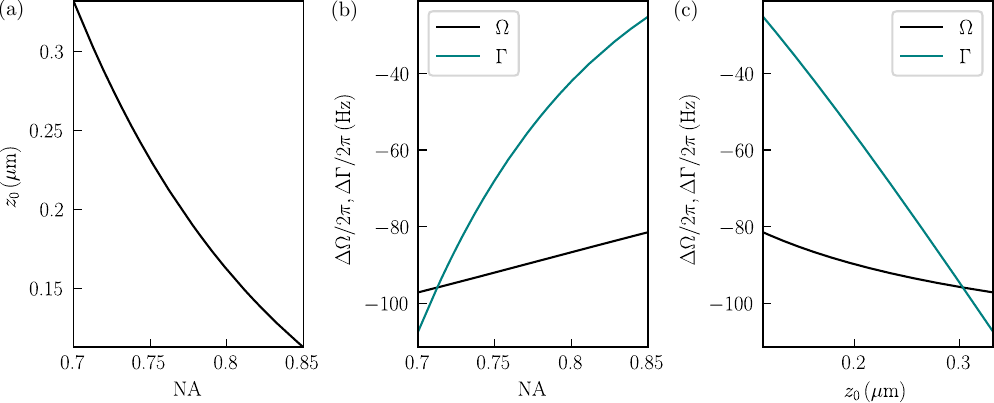}
    \caption{(a) Trapping position, $z_0$, for a \(75\) nm-radius particle levitated using a \(1064\) nm-wavelength and 40 mW-power laser when the objective's $\text{NA}$ is varied. 
    (b) Difference in trapping frequency, $\Delta \Omega$, and photon-recoil heating rate, $\Delta \Gamma$, between a trap at focus and at $z_0$, as a function of $\text{NA}$. 
    (c) Difference in trapping frequency and photon-recoil heating rate as a function of $z_0$.}
    \label{fig:eqpos}
\end{figure}
It is then fairly easy to see that, in the absence of phase modulation (i.e., $g(\theta)=0$), it is impossible to trap at focus (i.e., $z_0 = 0$), as often assumed in the literature \cite{rashid2017wigner}. 
Indeed, \(S_n\) vanishes at \(z=0\) for all \(n\in\mathbb{N}\) and, as the optical theorem imposes that $\alpha''\ne0$, stable trapping would require either $C_0=0$ or $C_1=0$, which cannot be fulfilled. 
The non-zero trapping position ensures that the particle experiences an effective force landscape that is different from that of the ideal strongly focused beam \cite{kleine_wavefront_2025}. 
As an illustration, in Figure \ref{fig:eqpos}, we plot the influence of $z_0$ onto optomechanical parameters.
Figure \ref{fig:eqpos}(a) displays the evolution of $z_0$ for a uniform wavefront, $g(\theta)=0$, when varying the numerical aperture, $\text{NA}$.
This curve emphasizes that a larger $\text{NA}$ tends to reduce the scattering force and brings the particle closer to focus.
When varying the $\text{NA}$, Figure \ref{fig:eqpos}(b) plots in black (respectively green), the difference $\Delta \Omega$ (respectively $\Delta \Gamma$) between the mechanical frequencies (respectively recoil-heating rates) computed for a trapping on focus ($z=0$) and at $z_0$ (i.e., fulfilling eq. \eqref{eq_position}).
Finally, Figure \ref{fig:eqpos}(c) combines panels (a) and (b) to relate $z_0$ with the mechanical frequency and recoil-heating rate discrepancies in black and green, respectively.

\newpage
\section{Numerical methods}
Now, we discuss the numerical method employed to compute the various parameters, as well as to implement the optimization procedure.

\subsection{Wavefront's modulation basis}
We expand our wavefronts as linear combinations of Zernike polynomials:
\begin{equation}
\label{phasemask}
    g(\theta) = \sum_1^{2N} a_nZ_n^0(\sin\theta),
\end{equation}
where \(\rho = \sin\theta\) and we restrict ourselves to axial-symmetric patterns (\(m = 0\)). 
Throughout this work, we use \(N = 11\) and exclude the constant \(n = 0\) phase pattern which has no impact. 
This choice corresponds to a trade-off between the accessible degrees of freedom within the optical field and computational speed. 
Axial-symmetric polynomials read:
\begin{equation}
\label{zernike}
    Z_n^0(\rho) = \sum_{k=0}^{n/2}\frac{(-1)^k(n-k)!}{k![(\frac{n}{2}-k)!]^2}\rho^{n-2k},
\end{equation}
for \(n\) even. 
The Zernike polynomials are computed by recurrence, following a method introduced by Prata and Rusch \cite{Prata:89}:
\begin{equation}
\label{recurence}
    \begin{split}
        &Z_0^0(\rho) = 1, \\
        &Z_2^0(\rho) = 2\rho^2-1,\\
        &Z_{n+2}^0(\rho) = \frac{1}{n+2}\left(4(n+1)\rho^2- 2n -2\right)Z_n^0(\rho) -nZ_{n-2}^0(\rho)
    \end{split}
\end{equation}

\begin{figure}[h]
\label{numerrors}
    \centering
    \includegraphics[width=\linewidth]{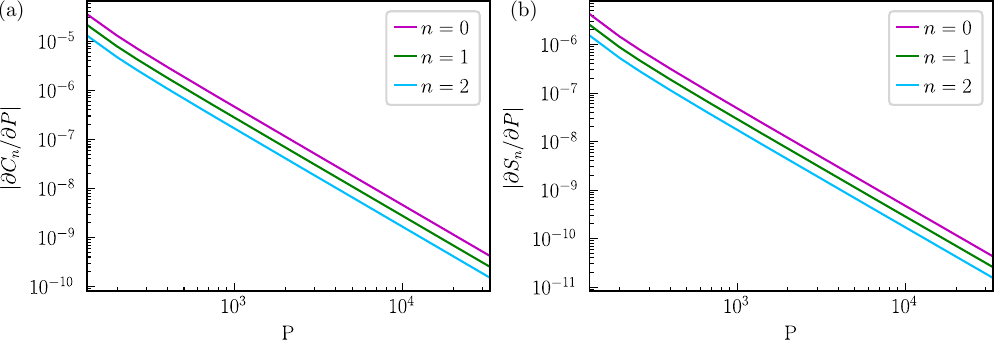}
    \caption{(a, b) Variations of the coefficients \(C_n\) and \(S_n\), respectively, when varying the number of points \(P\) used to sample the wavefront along the angular direction $\theta$. 
    The coefficient are here caculated for a uniform pattern and using the parameters of Figure \ref{fig:Stiffplot}.
    Typical values of $C_n$ and $S_n$ are usually around \(10^{-1}\) but they can go as low as \(10^{-4}\) in some cases. 
    This plot highlights the accuracy of our sampling.}
    \label{fig:Integration}
\end{figure}

Numerically, to compute the coefficients $C_n$ and $S_n$ (providing access to all physical quantities), the angular coordinate $\theta$ of the wavefront $g(\theta)$ is sampled uniformly from 0 to $\theta_{max} = \arcsin{(\text{NA})}$ using $P$=10000 points. 
To show that our sampling is accurate enough, in Figure \ref{fig:Integration}, we plot the evolution of different $C_n,S_n$ when varying the number of sampling point $P$.

\subsection{Optimization procedure}
\label{section:routine}
Our numerical routine relies on an optimization procedure performed using scipy and the implementation of the trust-region-interior-point method for constrained optimization \cite{Byrd1999AnIP}. 
Numerically, each iteration is characterized by a wavefront $g(\theta)$ corresponding to a linear combination of Zernike polyniomials, from which we implement a cost function  \(F\) that reads:
\begin{equation}
\label{Objectives}
    F(g) = \frac{\Gamma(g)}{\Gamma_0}\frac{\Omega_0}{\Omega_z(g)},
\end{equation}
in which $\Gamma(g)$ and $\Omega_z(g)$ (respectively $\Gamma_0$ and $\Omega_0$) stand for the recoil-heating rate and the mechanical frequency obtained with the wavefront $g$ (respectively for the uniform wavefront $g=0$).
Therefore, we ensure that the cost function provides an enhancement (or a reduction) with respect to the uniform case while being independent of the optical power.  
The optimization can either maximized of minimized recoil heating by changing the sign of the cost function. 

To ensure stable trapping, our code must fulfill the requirement introduced in eq. \eqref{eq_position}. 
Numerically, this translates into performing this optimization under an inequality as well as an equality constraint, which read respectively
\begin{equation}
    \frac{\Omega_z(g)^2}{\Omega_0^2} > c  > 0 \quad \& \quad F_{tot}(z_0)=0
    \label{constraints}
\end{equation}
The first condition, governed by the parameter $c$, enforces that the trap remains stable under realistic conditions by limiting any reduction of the optimized stiffness, $\propto \Omega(g)^2$, with respect to the unmodulated one, $\propto \Omega_0^2$. 
Specifically, the parameter \(c\) controls the strictness of the condition and corresponds to the increase in optical power an operator would have to provide to bring the frequency, $\Omega_z$, back to its initial value, $\Omega_0$. 
The second constraint ensures that the particle is trapped at \(z_0\) and the wavefront $g(\theta)$ fulfills $F_{tot}(z_0)=0$.

\begin{figure}[h]
    \centering
    \includegraphics[width=1\linewidth]{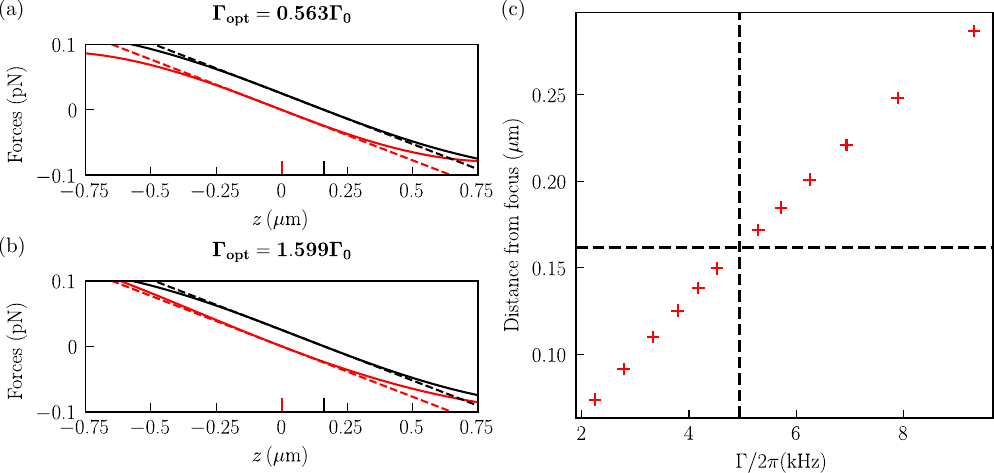}
    \caption{(a, b) Total optical forces before (thick black) and after (thick red) a minimization (a) and a maniximization (b) procedure applied to the configuration considered in Figure \ref{fig:Stiffplot}.
    In both panels the black and red ticks show trapping position before and after optimization, while the dashed lines are the tangents to the optical force at equilibrium, computed using eq. \eqref{Stiffness} (slope corresponding to \(-\kappa_z\)). 
    The results in term of recoil-heating rates are shown at the top of each panel. 
    (c) Distance between the trapping location and the focal spot for different optimizations obtained by adjusting the constraint parameter \(c\) in eq. \eqref{Objectives}. 
    For each optimization, a red cross marks the distance obtained as a function of its corresponding recoil-heating rate.  
    The black dashed lines show the distance from focus and recoil-heating rate before optimization. 
    For each panel, the simulations relied on the parameters used in Figure \ref{fig:Stiffplot}.}
    \label{fig:Optimisations}
\end{figure}

Figures \ref{fig:Optimisations}(a) and (b) display examples of a minimization and a maximization, respectively, using the parameters introduced in Figure \ref{fig:Stiffplot}. 
The variations achieved by both optimizations are respectively \(0.563\) and \(1.599\) after the laser intensity has been adjusted to bring $\Omega_z$ (and thus the stiffness) back to its unmodulated value, $\Omega_0$. 
There, the thick black (respectively thick red) curve plots the total force landscape measured for a uniform (respectively optimized) wavefront, while the dashed black (respectively red dashed) curve represents the tangent at the corresponding trapping location $z_0$ marked by a black tick (respectively red tick).
In each case, the uniform and optimized tangents are parallel, which emphasizes that the stiffness remains constant once the power has been increased by a factor \(\Omega_0^2/\Omega(g)^2\). 

In Figure \ref{fig:Optimisations} (c), the optimization procedure is reproduced for different values of the constraint parameter $c$ (see discussion in section \ref{section:Modulation}).
The horizontal and vertical black dashed lines mark the recoil-heating rate, $\Gamma$, and the trapping location, $z_0$, computed for a uniform pattern, respectively. 
The red crosses to the right (respectively left) of the vertical dashed line indicate maximization (respectively minimization) of $\Gamma$.
For each maximization or minimization, we compute the distance between the trapping location $z_0$ and the focal point, which does not necessarily coincide with the origin for a non-uniform wavefront $g(\theta)$ and corresponds to a zero of the gradient force (i.e., $S_0C_1-C_0S_1=0$).
This plot emphasizes the importance of the distance between the actual trapping position and the effective focus point. 
We notice that this distance increases when maximizing recoil heating, and decreases when minimizing it. 
This phenomenon highlights that the distance between the trap and the focus (which is \(z_0\) in the uniform case) constitutes an important degree of freedom, as it controls the ratio between gradient and scattering effects \cite{kleine_wavefront_2025}.

\subsection{Maxwell-stress-tensor simulations}
\begin{figure}[h]
    \centering
    \includegraphics[width=0.9\linewidth]{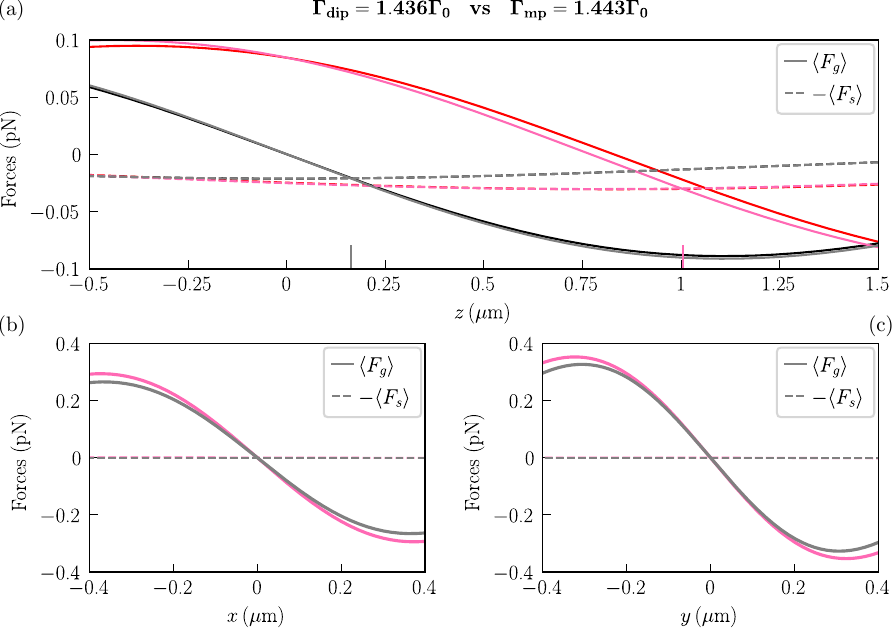}
    \caption{(a) Force landscapes (i.e., along \(z\)) computed using our full-wave method for a uniform pattern (gray) and an optimized pattern (pink). 
    The landscapes computed using our dipolar model are reported in black and red.
    For each color, the thick and dashed curve, correspond to the gradient force, $\langle F_g\rangle$, and the opposite of the scattering force $-\langle F_s \rangle$, respectively (see legend). 
    (b,c) Transverse force landscapes (i.e., along \(x\) and \(y\), respectively) computed through our full-wave method. The gray (respectively pink) curves stands for the forces computed for a uniform and optimized pattern, respectively. }
    \label{fig:Joel}
\end{figure}

To confirm the validity of our theoretical approach, we compared our results with a full-wave model providing the exact optical force landscapes along $x$, $y$ and $z$ (Figure. \ref{fig:Joel}). 
This model corresponds to a modified version of the Debye’s integral \cite{Novotny_Hecht_2012_2}, in which the wavefront produced by the SLM is modeled through a thin-lens apodization function describing a non-uniform phase distribution. 
The scattered field is obtained using the Generalized Lorentz Mie Theory (GLMT), as an expansion on vectorial spherical harmonics. 
Integrating the Maxwell Stress Tensor over a spherical shell surrounding the nanoparticle, we obtain the exact expression of the total optical force landscape. 
A computation of the scattered-field distributions for different locations of the nanoparticle provides access to the distribution of inelastic scattering, from which the recoil-heating rate is deduced.

First, we perform a maximization following the routine introduced in section \ref{section:routine}.
Figure \ref{fig:Joel}(a) displays in thick black and thick red the gradient forces, $\langle F_g \rangle$, computed for a uniform and optimized wavefront, respectively, while the dashed-black and dashed-red curves provide the opposite of the mean scattering force, $-\langle F_s \rangle$, computed for a uniform and and an optimized wavefront, respectively.
We report a final recoil heating $\Gamma_{dip} = 1.436 \Gamma_0$ with our dipolar approach, in which $\Gamma_0$ stands for the rate computed for a uniform wavefront.  
Then, the optimized wavefront obtained through this approach is inserted into our full-wave model.
Figure \ref{fig:Joel}(a) displays in thick gray and thick pink the gradient forces, $\langle F_g \rangle$, computed for a uniform and an optimized wavefront, respectively, while the dashed-gray and dashed-pink curves provide the opposite of the mean scattering force, $-\langle F_s \rangle$, computed for a uniform and an optimized wavefront, respectively.
We report a final recoil heating $\Gamma_{mp} = 1.443 \Gamma_0$ with this full-wave approach.
This minor difference with respect to the dipolar case can be explained by the contributions of other multipoles and in particular the magnetic dipole.
Overall, the agreement between the different landscapes as well as the optimized recoil-heating rates confirm the validity of our model.

At last, as showed in Figures \ref{fig:Optimisations}(b) and (c), this model also confirms that our trapping is stable along the transverse directions $x$ and $y$.

\clearpage
\section{Dynamic recoil-heating modulation}
\label{section:Modulation}

\begin{figure}[h]
    \centering
    \includegraphics[width=0.75\linewidth]{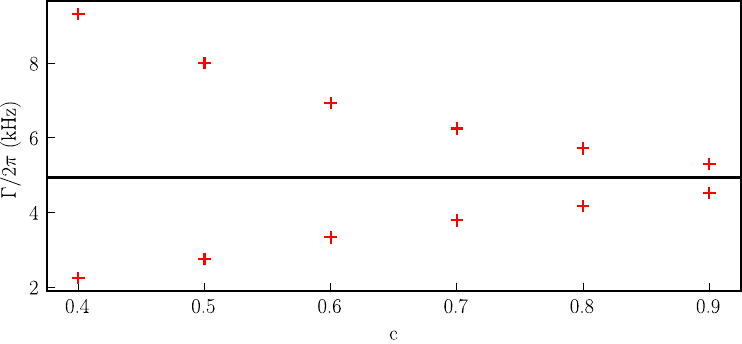}
    \caption{
    The minimizations and maximization reported in Figure \ref{fig:Optimisations}(a) and (b) are reproduced for different values of the parameter constraint $c$ introduced in eq. \eqref{constraints}.
    For each value of $c$, two red crosses indicate respectively the maximized and minimized recoil-heating rates, which lay respectively above and below the unmodulated value marked by the black line. 
    The physical parameters used for the optimizations are the same as those used in Figure \ref{fig:Stiffplot} and \ref{fig:Optimisations}.}
    \label{fig:bykappa}
\end{figure}

\begin{figure}[h]
    \centering
    \includegraphics[width=0.8\linewidth]{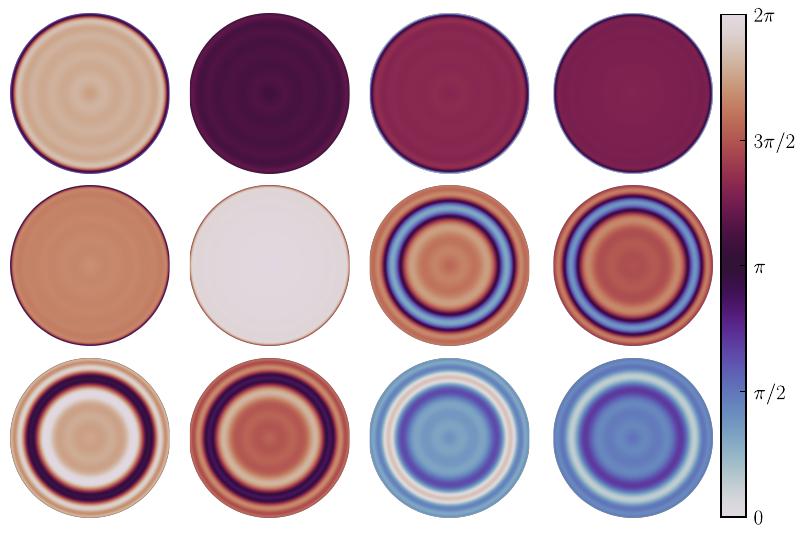}
    \caption{Series of phase patterns \(g(\theta)\) use to enforce the recoil-heating modulation displayed in Figure 3 of the main text.}
    \label{fig:Collection}
\end{figure}

At last, the optimization constraints of eq. \eqref{constraints} can be adapted to provide a smooth and dynamic modulation of the recoil-heating rate.
In particular, in Figure \ref{fig:bykappa}, we tune the parameter $c$, while performing both a maximization and a minimization for each value.
We clearly observe that, when varying this parameter, larger recoil-heating modifications can be achieved.
By switching  between the different wavefronts obtained while varying $c$, one can perform a smooth modulation.
As an illustration, Figure \ref{fig:Collection} provides the 12-pattern sequence used to perform the modulation displayed in Figure 3 of the main text.   
Of course, the parameter $c$ can be adjusted to provide any type of modulation one could envision.

Finally, we stress that switching between different wavefronts can be achieved while ensuring a constant trapping location, $z_0$.
Indeed, defocus can be added to a given phase pattern to translate $z_0$ to any location.
Specifically, in eq. \eqref{CS}, replacing $g(\theta)$ by $g(\theta) - 2\pi /\lambda \Delta z \cos{\theta}$ translates the trapping location to $z_0 + \Delta z$, while leaving unaffected the coefficients $C_n$ and $S_n$ that set all relevant optomechanical parameters.

\newpage
\printbibliography

\end{document}